\documentstyle[preprint,aps,epsfig]{revtex}

\begin{document}
\title{Electron-Phonon Interaction and Raman Linewidth in Superconducting
Fullerides.}
\author{V.L. Aksenov, V.V. Kabanov}
\address{Frank Laboratory of Neutron Physics, JINR, 141980 Dubna, Russia}
\maketitle

\begin{abstract}
We propose a microscopic theory of interaction of long wave molecular
phonons with electrons in fullerides in the presence of disorder. Phonon
relaxation rate and frequency renormalization are discussed. Finite
electronic bandwidth reduces phonon relaxation rate at $q=0$.
Electron-phonon coupling constants with molecular modes in fullerides are
estimated. The results are in good agreement with photoemission experiments.
\end{abstract}


PACS numbers 79.60.Bm, 63.20.-e, 71.38.+i

\narrowtext

\section{Introduction.}

\noindent Superconducting fullerides are a new type of materials, where the
electronic bandwidth is of the same order as the frequencies of
intramolecular modes. \cite{pic,gelf}. Nonadiabaticity of electrons measured
by the ratio of characteristic phonon frequency $\omega$ to the Fermi energy 
$E_{F}$ is not small. The phonon frequencies are high $\omega \leq 0.2$ $eV$
and bare Fermi energy is low $E_{F} \leq 0.2$ $eV$\cite{pic}.

In the past years several different calculations of the electron-phonon
coupling constants have been reported for fullerides\cite
{pic1,antr,varma,shlu,faul}. Some of them yield the strongest coupling with
the high frequency $H_{g}$ modes with a moderate electron-phonon coupling, $%
\lambda \leq 0.5$. on the other hand picket $et$ $al$\cite{pic1} predicted
the strongest coupling with the high frequency $A_g(2)$ mode and $\lambda
\sim 3$. Similar conclusion has been reached in Ref.\cite{stolh}. The
difference in calculated coupling constants is quite remarkable, and may
result in a qualitatively different understanding of the nature of
superconductivity of fullerides. Therefore, the experimental determination
of $\lambda$ is required\cite{kuz1,kuz2,gun,alekab}.

Recently Raman spectra for metallic fullerides at low temperature has been
reported in Ref.\cite{kuz1,kuz2}. The linewidth have been analyzed using
Allen's formula for the decay rate of the phonon into electron-hole pair $%
\gamma$ averaged over all phonon momenta\cite{allen}: 
\begin{equation}
\bar{\gamma}=\frac {\pi N(0)\lambda \omega^{2}}{2\kappa},
\end{equation}
$N(0)$ is the density of states on the Fermi level, $\lambda$ is
electron-phonon coupling constant, $\kappa$ is degeneracy of the phonon mode.

It is well known that the phonon lifetime is determined by the parameter $%
qv_{F}/\omega $, $v_{F}$ is Fermi velocity, and $\omega $ is the frequency
of optical phonon. It means that Allen's formula for the phonon linewidth
does not work for optical phonons in $q\approx 0$ limit. In the clean
single-band system it is not possible for any optical $q\approx 0$ phonon to
decay into electronic excitations because of the conservation of the
momentum and the energy. $\Im \Pi (q,\omega )=0$ for $qv_{F}\ll \omega $ 
\cite{schr,ippat}. This result is based on the Ward identity and is
independent on vertex corrections (Eqs. (4.20) and (4.21) of the Ref. \cite
{schr}). Moreover high frequency phonon can not decay into electron-hole
pairs if $\omega \geq w$, $w$ is electronic half bandwidth.

It should be pointed out some special comments concerning the $pentagonal$ $%
pinch$ $A_{g}(2)$ mode. It shows only a little broadening with doping. The
authors in Refs.\cite{kuz1,kuz2,gun} conclude that it is the manifestation
of the weak coupling with this mode. As it has been mentioned by Gelfand\cite
{gelf} a $q=0$ $A_{g}$ mode shifts all the energy levels on $all$ molecules
in the solid by the same amount, and therefore leads to only diagonal
elements between band states for the deformation potential. The $q=0$ $A_{g}$
modes are thus uncapable of decaying into an electron-hole pair, no matter
how strong the electron-phonon coupling is.

Thus, because Fermi velocity $v_{F}$ is small and the frequency of
intramolecular modes are high, the ratio $v_{F}/\omega$ is small and formula
for the phonon lifetime in adiabatic limit $\omega \rightarrow 0$ does not
work. The finite contribution to the phonon lifetime for $\omega \rightarrow
0$ appears due to impurity scattering and orientational disorder\cite
{schluter1} and the violation of the conservation of the momentum.

In this paper we analyze the phonon relaxation rate and the renormalization
of the phonon frequency for $q=0$ due to electron-phonon interaction in the
presence of disorder and taking into account finite electronic bandwidth. We
take into account the effect of disorder in terms of relaxation time $\tau$
and adopt Fermi-liquid description $E_{F}\tau \gg 1$.

\section{Effect of disorder.}

\noindent We describe electron-phonon interaction in fullerides by the
standard hamiltonian \cite{lanoo}. It describes the interaction of $t_{1u}$
electrons with $A_{g}$ and $H_{g}$ intramolecular modes: 
\begin{equation}
H=\sum_{k,\sigma,i} \epsilon_{k}c^{\dagger}_{k,\sigma,i} c_{k,\sigma,i}
+\sum_{k,q,\sigma,i}g_{k} c^{\dagger}_{k,\sigma,i}c_{k+q,\sigma,i}
(b^{\dagger}_{q}+b_{-q})+\sum_{q}\omega b^{\dagger}_{q}b_{q},
\end{equation}
where the first term is the kinetic energy of the electrons in threefold
degenerate $t_{1u}$ band, $c^{\dagger}_{k,\sigma,i}$ is the creation
operator of the electron with momentum $k$, spin $\sigma$ and orbital index $%
i$($i = 1,2,3$), $b^{\dagger}_{q}$ is the creation operator of the phonon
with momentum $q$. Here we take into account momentum dependence of the
coupling constant explicitly. For intramolecular modes this dependence is
weak, but as we discuss later this dependence is responsible for the finite
contribution of the electron-phonon coupling to the phonon relaxation rate
at $q=0$.

Note that the fine structure of $H_{g}$ phonons is usually neglected for the
analysis of the relaxation rate with Allen's formula\cite{kuz2}. We also
neglect strong degeneracy of $H_{g}$ modes. This assumption is quite
reasonable if electronic relaxation time is large $\omega \tau >>1$ and if
crystal field effects are strong and split of the fivefold degenerate modes
is strong\cite{kuz2}. It is clear because the nodiagonal elements of the
electronic Green's function appear only due to impurity scaterring and are
small if $\omega \tau >>1$. It is important, that the interaction constant
with single $H_{g}$ submode is strongly momentum dependent.

Phonon relaxation rate and frequency renormalization are determined by the
real and imaginary parts of the polarization : 
\begin{equation}
\Pi (q=0,\omega )=i\int \Gamma (k,\omega ^{^{\prime }}+\omega /2,\omega
^{^{\prime }}-\omega /2)g_{k}G(k,\omega ^{^{\prime }}+\omega /2)G(k,\omega
^{^{\prime }}-\omega /2)\frac{d^{3}kd\omega }{(2\pi )^{4}}
\end{equation}
The equation for the vertex has the form\cite{agd} (Fig.1): 
\begin{eqnarray}
\Gamma (k,\omega ^{^{\prime }}+\omega /2,\omega ^{^{\prime }}-\omega /2)
&=&g_{k}+n_{im}/(2\pi )^{3}\int |u(p-k)|^{2}G(p,\omega ^{^{\prime }}+\omega
/2) \\
&&G(p,\omega ^{^{\prime }}-\omega /2)\Gamma (p,\omega ^{^{\prime }}+\omega
/2,\omega ^{^{\prime }}-\omega /2)d^{3}p  \nonumber
\end{eqnarray}
where $u(p-k)$ is the potential of the single impurity, $G(k,\omega
)=1/(\omega -\xi -\Sigma (\omega ))$ is electronic Green function, averaged
over impurity\cite{agd}, $\Sigma (\omega )\simeq -i\frac{\omega }{2|\omega
|\tau }$, $\tau $ is electronic relaxation time, $n_{im}$ is concentration
of impurities.

We define the function: 
\begin{equation}
P(k,\omega^{^{\prime}}+\omega/2,\omega^{^{\prime}}-\omega/2)=
\Gamma(k,\omega^{^{\prime}}+\omega/2,\omega^{^{\prime}}-\omega/2)
G(k,\omega^{^{\prime}}+\omega/2) G(k,\omega^{^{\prime}}-\omega/2).
\end{equation}
This function satisfies the equation: 
\begin{eqnarray}
P(k,\omega^{^{\prime}}+\omega/2,\omega^{^{\prime}}-\omega/2)=
G(k,\omega^{^{\prime}}+\omega/2) G(k,\omega^{^{\prime}}-\omega/2)( g_{k}+ \\
n_{im}/(2\pi)^{3}\int |u(p-k)|^{2}
P(p,\omega^{^{\prime}}+\omega/2,\omega^{^{\prime}}-\omega/2) d^{3}p) 
\nonumber
\end{eqnarray}
Main contribution to the integrals appears from the momenta near the Fermi
surface $k \sim k_{F}$ and we can expand $g_{k}$ and $|u(k-p|^{2}$ in
spherical harmonics $\phi_{L}(k)$ on the Fermi surface \cite{zawa}: 
\begin{equation}
g_{k}=\sum_{L}g_{L}\phi_{L}(k)
\end{equation}
\begin{equation}
|u(p-k)|^2= \sum_{L,L^{^{\prime}}}\phi_{L}(k)
\Gamma_{L,L^{^{\prime}}}\phi_{L^{^{\prime}}}(p)^{*}
\end{equation}
For the sake of simplicity we suppose that $\Gamma_{L,L^{^{\prime}}} =
\delta_{L,L^{^{\prime}}}\Gamma_{L}$. The equations for the relaxation times
have the form $1/\tau = 2\pi N(0) n_{im}\Gamma_{0}$, $1/\tau_{L} = 2\pi N(0)
n_{im}\Gamma_{L}$, where $N(0)$ is the density of state on the Fermi level.
Note that $g_{L=0} \gg g_{L \neq 0}$ for $A_{g}$ modes. On the other hand
for fivefold degenerate $H_{g}$ modes we expect strong $k$ dependence of the
coupling constant.

We define the set of functions $\Lambda_{L}(\omega^{^{\prime}},\omega)$: 
\begin{equation}
\sum_{L}g_{L}\phi_{L}(k)\Lambda_{L}(\omega^{^{\prime}},\omega)=
n_{im}/(2\pi)^{3}
\int|u(k-p^{^{\prime}})|^{2}P(p^{^{\prime}},\omega^{^{\prime}}+\omega/2,%
\omega^{^{\prime}}-\omega/2) d^{3}p,
\end{equation}
and derive the equation for $\Lambda_{L}(\omega^{^{\prime}},\omega)$: 
\begin{eqnarray}
\sum_{L}g_{L}\phi_{L}(l)\Lambda_{L}(\omega^{^{\prime}},
\omega)=n_{im}/(2\pi)^{3}
\sum_{M}(1+\Lambda_{M}(\omega^{^{\prime}},\omega)\int|u(k-p)|^{2}g_{M}
\phi_{M}(p) \\
G(p,\omega^{^{\prime}}+\omega/2)G(p,\omega^{^{\prime}}-\omega/2)d^{3}p. 
\nonumber
\end{eqnarray}
Integrating out the angles in Eq.(10) and taking into account Eq.(8) we
obtain: 
\begin{eqnarray}
\Lambda_{L}(\omega^{^{\prime}},\omega)=\hspace{1cm} i/\tau_{L} \frac{1}{%
\omega+i/\tau_{L}^{*}} \hspace{1cm} |\omega^{^{\prime}}| < |\omega| \\
0 \hspace{2.5cm} |\omega^{^{\prime}}| > |\omega|  \nonumber
\end{eqnarray}
where $1/\tau_{L}^{*} = 1/\tau - 1/\tau_{L}$

Note that for $L=0$ $\Lambda_{0}(\omega^{^{\prime}},\omega)=i/\tau\omega$.
The largest term in the expansion of the coupling constant $g_{0}$ does not
contribute to the $q=0$ phonon relaxation rate. Substituting Eq.(5) to
Eq.(3) and taking into account Eqs.(9),(6) and (11) we obtain: 
\begin{eqnarray}
\Pi(0,\omega) = -2 i \sum_{L \neq 0} \frac {g_{L}^{2} N(0)/\tau_{L}^{*}}{%
\omega+i/\tau_{L}^{*}}.
\end{eqnarray}
Here we take into account that $\int
d\omega^{^{\prime}}(\Sigma(\omega+\omega^{^{\prime}}/2)-\Sigma(\omega-%
\omega^{^{\prime}}/2)) = 0$.

As a result we obtain the formula for the phonon relaxation rate $%
\gamma(\omega)$: 
\begin{equation}
\gamma(\omega) = - \Im\Pi(0,\omega)=2\sum_{L \neq 0}\frac{%
g_{L}^{2}N(0)\omega \tau_{L}^{*}}{\omega^{2}\tau_{L}^{*2}+1}
\end{equation}
It follows from the Eq.(13) phonon relaxation rate at $q \rightarrow 0$ is
determined by the parameter $<g_{k}^{2}>-<g_{k}>^{2}$, where $<..>$ is
average over Fermi surface. This formula is strongly different from Allen's
formula\cite{allen}. ($i$) Phonon relaxation rate is proportional to the
averaged over Fermi surface $k$-dependent component of the electron-phonon
coupling constant. Phonon relaxation rate due to electron-phonon coupling is
equal to zero if coupling constant is independent of the electronic momentum 
$k$. ($ii$) Phonon relaxation rate is proportional to the impurity
scattering relaxation rate of electrons at low temperatures $1/\tau^{*}$.
Therefore, momentum dependence of the electron-phonon interaction is
responsible for the finite Raman linewidth.

It should be pointed out, that similar formula for the relaxation rate of
the optical phonons in metals was derived from kinetic equation in Ref.\cite
{misc} and Green's function technique\cite{kost}. Note that formula (13) is
different from that derived in Ref.\cite{misc}. New term proportional to $%
\Lambda (\omega ^{^{\prime }},\omega )$ appears in the equation for $\Pi
(\omega )$ due to correct average of the vertex over impurities. Neglecting
this term one can derive the same formula for relaxation rate as Eq.(18) of
Ref.\cite{misc}.

Extensive numerical calculations of the phonon lifetime, using spherically
symmetrical coupling have been performed in Ref.\cite{des}. It has been
shown that diagonal component of the polarization is site dependent in
disordered phase. This fact is in agreement with formula (13). Because $%
H_{g} $ modes are not spherically symmetrical the interaction with the five
split submodes will have large $L \ne 0$ harmonics on the Fermi surface even
in the case of spherically symmetrical bare interaction.

\section{Bandwidth effect.}

\noindent In superconducting fullerides there are a number of molecular
modes with the frequencies of the order of bare bandwidths. These are $%
pentagonal$ $pinch$ mode $A_{g}(2)$ $\omega \simeq 1500 cm^{-1}$ and four $%
H_{g}$ modes with $\omega \simeq 1200-1600cm^{-1}$. Because of conservation
of energy these modes cannot decay into electron-hole pair in the clean
system. Note that in the limit of $w \ll \omega$ phonon relaxation rate is
equal to 0 in the lowest order in coupling constant.

We use Eqs. (3) and (4) for the polarization and lorenzian form of the
density of states to take into account the finite bandwidth: 
\begin{equation}
N(\xi) = \frac {2\nu}{\pi} \frac {w}{\xi^{2}+w^{2}}
\end{equation}
where $w$ is effective half bandwidth, $\nu$ is orbital degeneracy. For the $%
t_{1u}$ band $\nu = 3$. Using Eq.(14) we can derive the equation for
electronic self-energy averaged over impurities in ladder approximation\cite
{aek}: 
\begin{equation}
\Sigma(\omega) = xw^{2} \frac {1}{\omega+iw \omega/|\omega|)- \Sigma(\omega)}
\end{equation}
where $x = \nu\Gamma_{0}n_{im}/w^{2} = 1/2\tau w$ is dimensionless
concentration of impurities.

Integrating out the angle in Eq.(10) and taking into account Eq.(8) we
obtain the formula for $\Lambda_{L}(\omega^{^{\prime}},\omega)$: 
\begin{equation}
\Lambda_{L}(\omega^{^{\prime}},\omega) = \frac {x_{L}}{x} \frac {%
\Sigma(\omega-\omega^{^{\prime}}/2)-\Sigma(\omega+\omega^{^{\prime}}/2)} {%
\omega+\frac {x-x_{L}}{x} (\Sigma(\omega-\omega^{^{\prime}}/2)-\Sigma(%
\omega+\omega^{^{\prime}}/2)}
\end{equation}
where $x_{L} = \nu\Gamma_{L}n_{im}/w^{2} = 1/2\tau_{L} w$. We have used here
integral equation for the electronic self-energy in ladder approximation\cite
{agd,aek}. Note, that Eq.(16) is equivalent to the Eq.(11) if $%
\Sigma(\omega) =-\frac {i \omega}{2|\omega|\tau}$. Equation for the $L$
component of the polarization has the form: 
\begin{equation}
\Pi_{L}(\omega) = \frac {-2 i g_{L}^{2}\nu}{\pi x w} \int dy \frac {%
\Sigma(y-\omega/2)-\Sigma(y+\omega/2)} {\omega+\frac {x-x_{L}}{x}
(\Sigma(y-\omega/2)-\Sigma(y+\omega/2)}
\end{equation}
Taking into account that $E_{F}\tau \simeq w\tau \gg 1$ we obtain: 
\begin{equation}
\Sigma(\omega) = x w^{2} \frac {1}{\omega+iw \omega/|\omega|}
\end{equation}
Substituting Eq.(18) into Eq.(17) and integrating out $y$ we derive the
formulae for imaginary and real parts of the polarization: 
\begin{equation}
\Re\Pi(\omega) = \sum_{L}\frac {2g_{L}^{2}N(0)}{w(\omega/w)^{2} \tau_{L}^{*}}
(\frac {\omega\ln{(1+(\omega/w)^{2})}-4w\arctan{(\omega/w)}} {%
\omega((\omega/w)^{2}+4)}+1)
\end{equation}

\begin{equation}
\gamma(\omega) = -\Im\Pi(\omega) = \sum_{L}\frac {4g_{L}^{2}N(0)}{%
w(\omega/w)^{3}((\omega/w)^{2}+4)\tau_{L}^{*}} (\ln{(1+(\omega/w)^{2})}%
+\omega\arctan{(\omega/w)/w})
\end{equation}
Eq.(20) reduces to Eq.(13) in the large bandwidth limit $\omega/w \ll 1$. In
the opposite limit $\omega/w \gg 1$ the relaxation rate is strongly reduced: 
\begin{equation}
\gamma(\omega) = -\Im\Pi(\omega) = \sum_{L}\frac {2\pi g_{L}^{2}N(0)w^{3}}{%
\omega^{4}\tau_{L}^{*}}
\end{equation}

\section{Conclusion.}

\noindent In conclusion we analyze the experimental data on the Raman
scattering in fullreides\cite{kuz1,kuz2} using the correct formula for $q=0$
phonon relaxation rate. Unfortunately, direct estimate of the coupling
constant is practically impossible. It requires exact form of angular
dependence of electron-phonon coupling constant on the Fermi surface and
electronic relaxation rate $1/\tau$. However, if we assume that $g_{L}^{2}
\sim <g^{2}> \sim \lambda\omega/N(0)$ and use the value for $H_{g}(1)$ mode
from the photoemission experiments\cite{gun,alekab}, we can calculate
coupling constants for another 7 $H_{g}$ modes using the formula: 
\begin{equation}
\gamma_{i}/\gamma_{j} \sim \lambda_{i}/\lambda_{j}
\end{equation}

If we suppose that $\lambda_{1}/N(0) \simeq 0.02eV$\cite{gun,alekab} for $%
H_{g}(1)$ mode, we obtain the coupling constants $\lambda_{i}/N(0)$ for
other 7 $H_{g}$ modes (Table 1). Note that Eq.(22) is valid only for $H_{g}$
modes, because angular dependence of the coupling constant on the Fermi
surface for $A_{g}$ modes is strongly different from that for $H_{g}$ modes
and we do not expect the cancellation of angular factor in Eq.(13). From the
Table 1 we can conclude:

\begin{itemize}
\item  Using Eq.(22) and the experimental Raman linewidths we obtain
coupling constants for $H_{g}$ modes. They are in good agreement with
photoemission data. Note that Allen's formula underestimates the coupling
constants by the order of magnitude for the most of the $H_{g}$ modes.

\item  The difference in coupling constants for $H_{g}(2)$ and $H_{g}(3)$
modes is probably connected with the fact that in the analysis of
photoemission spectra of $C_{60}^{-}$ the interaction with $A_{g}(1)$ mode
has been neglected\cite{gun,alekab}.

\item  The difference in estimated constants for $H_{g}(7,8)$ modes is due
to frequency dependence of electronic relaxation time $\tau $. Because the
interaction with low frequency modes is quite strong we expect strong
frequency dependence of $\tau $ and Eq.(22) is not valid.

\item  Due to high symmetry of $A_{g}(1,2)$ modes angular dependence of the
coupling constants is weak and Eq.(22) does not work.
\end{itemize}

It should be pointed out that frequency renormalization of these modes is
not due to effects considered in the paper. Indeed, the downshift of $%
A_{g}(2)$ mode is about $6$ $cm^{-1}$ per elementary charge on the $C_{60}$.
If we suppose that this downshift is due to interaction of phonons with band
electrons one should expect the maximum of downshift near the half-filled
band ($x=3$) and the absence of the downshift for $x=6$. In an isolated
molecule there is also a frequency renormalization when molecule becomes
charged. Theoretical estimates of the frequency shift due to the charging of 
$C_{60}$ molecule are in a reasonable agreement with experiments\cite{fried}.

We have estimated coupling constants of the conducting electrons with the
molecular phonons in superconducting fullerides from Raman experiments. The
results are in good agreement with that obtained from photoemission
measurements. Note that these constants with proper account of polaron
effect lead to correct values of $T_{c}$, isotope effect and pressure
dependence of $T_{c}(P)$\cite{alekab}.

We highly appreciate enlightening discussions with A.S. Alexandrov, N.M.
Plakida, J. Annett, E.G. Maksimov, O.V. Dolgov and D. Mihailovic. One of us
(V.V.K.) thanks Russian Foundation for Basic Research (Grant 97-02-16705)
and Slovenian Ministry of Research and Technology for financial support and
Dragan Mihailovic for hospitality.

\newpage 
\begin{table}[tbp]
\caption{Coupling constants obtained from Raman measurements using Allen's
formula (AF), Eq.(13) and from photoemission experiments (PES).}
\begin{tabular}[t]{lllllll}
& $\omega$ & $\gamma$\cite{kuz2} & $\lambda/N(0)$(eV) & $\lambda/N(0)$(eV) & 
$\lambda/N(0)$(eV) & $\lambda/N(0)$(eV) \\ 
& (cm$^{-1}$) & (cm$^{-1}$) & AF\cite{kuz2} & Eq.(13) & PES\cite{gun} & PES%
\cite{alekab} \\ \hline
$H(1)$ & 270 & 20 & 0.048 & 0.020 & 0.019 & 0.020 \\ 
$H(2)$ & 432 & 21 & 0.020 & 0.021 & 0.040 & 0.038 \\ 
$H(3)$ & 709 & 8 & 0.002 & 0.008 & 0.013 & 0.019 \\ 
$H(4)$ & 773 & 10 & 0.003 & 0.010 & 0.018 & 0.018 \\ 
$H(5)$ & 1100 & 11 & 0.001 & 0.011 & 0.012 & 0.009 \\ 
$H(6)$ & 1248 & 10 & 0.001 & 0.010 & 0.005 & 0.001 \\ 
$H(7)$ & 1425 & 46 & 0.004 & 0.046 & 0.017 & 0.000 \\ 
$H(8)$ & 1572 & 42 & 0.003 & 0.042 & 0.023 & 0.000 \\ \hline
\end{tabular}
\end{table}
\vspace{1cm} \newpage 
\begin{figure}[h]
\caption{Equation for the vertex function $\Gamma(k,\omega^{^{\prime}}+%
\omega/2,\omega^{^{\prime}}-\omega/2)$}
\end{figure}

\end{document}